\documentclass[journal=jpclcd,manuscript-type=letter]{achemso}
\usepackage[utf8]{inputenc}
\usepackage{textgreek}

\usepackage{amssymb}
\usepackage{amsmath}
\usepackage{graphicx}
\usepackage{dcolumn}
\usepackage{xcolor}
\usepackage{fancyhdr}
\usepackage{chemformula}
\usepackage[colorlinks,allcolors=black,citecolor=blue,urlcolor=blue]{hyperref}
\usepackage{etoolbox}
\patchcmd{\tocentry}{\setlength{\fboxrule}{0.4pt}}{\setlength{\fboxrule}{0pt}}{}{}

\author{Samuel G.\ H.\ Brookes}
\affiliation{Yusuf Hamied Department of Chemistry, University of Cambridge, Lensfield Road, Cambridge, CB2 1EW, UK}
\alsoaffiliation{Cavendish Laboratory, Department of Physics, University of Cambridge, Cambridge, CB3 0US, UK}
\alsoaffiliation{Lennard--Jones Centre, University of Cambridge, Trinity Ln, Cambridge, CB2 1TN, UK}

\author{IniOluwa C.\ Popoola}
\affiliation{Yusuf Hamied Department of Chemistry, University of Cambridge, Lensfield Road, Cambridge, CB2 1EW, UK}

\author{Fabian Berger}
\affiliation{Yusuf Hamied Department of Chemistry, University of Cambridge, Lensfield Road, Cambridge, CB2 1EW, UK}
\alsoaffiliation{Lennard--Jones Centre, University of Cambridge, Trinity Ln, Cambridge, CB2 1TN, UK}
\alsoaffiliation{Max Planck Institute for Polymer Research, Ackermannweg 10, 55128 Mainz, Germany}

\author{Kara D.\ Fong}
\affiliation{Yusuf Hamied Department of Chemistry, University of Cambridge, Lensfield Road, Cambridge, CB2 1EW, UK}
\alsoaffiliation{Lennard--Jones Centre, University of Cambridge, Trinity Ln, Cambridge, CB2 1TN, UK}
\alsoaffiliation{Division of Chemistry and Chemical Engineering, California Institute of Technology, Pasadena, California 91125, USA}
\alsoaffiliation{Marcus Center for Theoretical Chemistry, California Institute of Technology, Pasadena, CA 91125, USA}

\author{Angelos Michaelides}
\email{am452@cam.ac.uk}
\affiliation{Yusuf Hamied Department of Chemistry, University of Cambridge, Lensfield Road, Cambridge, CB2 1EW, UK}
\alsoaffiliation{Lennard--Jones Centre, University of Cambridge, Trinity Ln, Cambridge, CB2 1TN, UK}

\author{Christoph Schran}
\email{cs2121@cam.ac.uk}
\affiliation{Cavendish Laboratory, Department of Physics, University of Cambridge, Cambridge, CB3 0US, UK}
\alsoaffiliation{Lennard--Jones Centre, University of Cambridge, Trinity Ln, Cambridge, CB2 1TN, UK}

\title{Breaking Water at Graphene Defects}

\begin{tocentry}
  \includegraphics[width=3.25in]{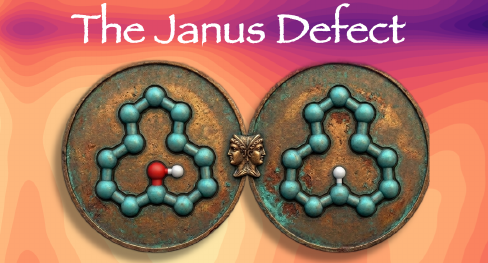}
\end{tocentry}

\begin{document}

\begin{abstract}
Water dissociation at solid surfaces underpins processes ranging from corrosion and catalysis to electrochemistry and photovoltaics.
Defects often serve as reactive sites for dissociation, yet how solvation influences water dissociation at such sites remains poorly understood.
Here, we use state-of-the-art machine-learned interatomic potentials to explore water dissociation at defective graphene-water interfaces.
We show that solvation qualitatively changes the reaction mechanism at a graphene single vacancy (SV), opening pathways that are absent for an isolated water molecule.
Whereas the gas-phase process proceeds via a single concerted channel, the solvated SV splits water through two competing pathways: a basic route forming SV-H and \ch{OH-}$_\text{(aq)}$, and an acidic route forming SV-OH and \ch{H3O+}$_\text{(aq)}$. 
These lower-barrier pathways produce distinct chemisorbed intermediates that enhance graphene-water adsorption.
Accordingly, even a simple carbon vacancy gives rise to unexpectedly rich interfacial chemistry, coupling surface chemistry to interfacial charge and wettability, with implications for carbon functionalization and nanofluidic transport.
\end{abstract}

\noindent The dissociation of water at a solid surface is a fundamental elementary reaction step in chemistry \cite{Carrasco2012,Olle2016}. 
In the gas phase, it takes 5 eV to cleave an O-H bond \cite{Ruscic2002}; however, when in contact with a surface, this process can proceed at a fraction of the energetic cost, with reported barriers often below 1 eV \cite{Phatak2009}.
Surface-mediated water dissociation underpins several key processes, including the Volmer step of alkaline hydrogen evolution \cite{Subbaraman2011},  the water-gas shift reaction \cite{Rodriguez2007}, and photocatalytic water splitting \cite{Fujishima1972,Maeda2006}.
Left unchecked, the same chemistry corrodes carbon electrodes in fuel cells and degrades metal and oxide surfaces exposed to aqueous environments \cite{Borup2007}.
Whether a given surface dissociates water, and by which mechanism, depends sensitively on the electronic structure of the surface, its local coordination and geometry, and the hydrogen-bonding environment provided by the surrounding water \cite{Henderson2002,Olle2016}.

There are a number of mechanisms by which water dissociates at a surface. 
On flat transition metals, water molecules can undergo direct dissociative chemisorption, colliding with the metal surface and cleaving an O-H bond to deposit H and OH fragments via a concerted mechanism \cite{Henderson2002}.
Solvation with other water molecules can facilitate this process \cite{Michaelides2003,Michaelides2004}.
When corrugations, steps, or kinks are introduced to the surface, metal-water dissociation can be further enhanced with binding to undercoordinated sites \cite{Desai2003,Donadio2012,Donadio2017}.
On metal oxide surfaces such as rutile \ch{TiO2}(110), oxygen vacancies can help promote dissociation via heterolytic mechanisms in which a proton transfers to a neighboring surface oxygen, yielding two surface hydroxyls per defect\cite{Schaub2001,Bikondoa2006}.
The surrounding solvent can also play an important role; in the case of pristine \ch{TiO2}, water dissociation can proceed through solvent-assisted proton transfer pathways that vary from one-step to two-step processes depending on the facet \cite{CalegariFreeEnergy,Zhuang2022,WenTiO2PNAS,ZengTiO2NatComm}.
Similar proton-transfer phenomena occur at solvated ZnO surfaces \cite{Tocci2014}.
Yet, while defect-mediated reactivity and solvent-assisted proton transfer are well-established for metals and metal oxides, how these two phenomena couple together for 2D materials such as graphene, \ch{MoS2}, or hBN remains far less understood \cite{ZhangMoS2EES,WanghBNJACS,ComtethBN,Tocci2016,Scalfi2023,Zhang2026}.

These gaps motivate two key open questions:
(i) how does solvation alter the mechanism and energetics of water dissociation for these 2D materials, and
(ii) how do the resulting chemisorbed species modify the structure and adsorption properties of the surrounding water?

\begin{figure*}[t!]
\centering
    \includegraphics[width=0.75\textwidth]{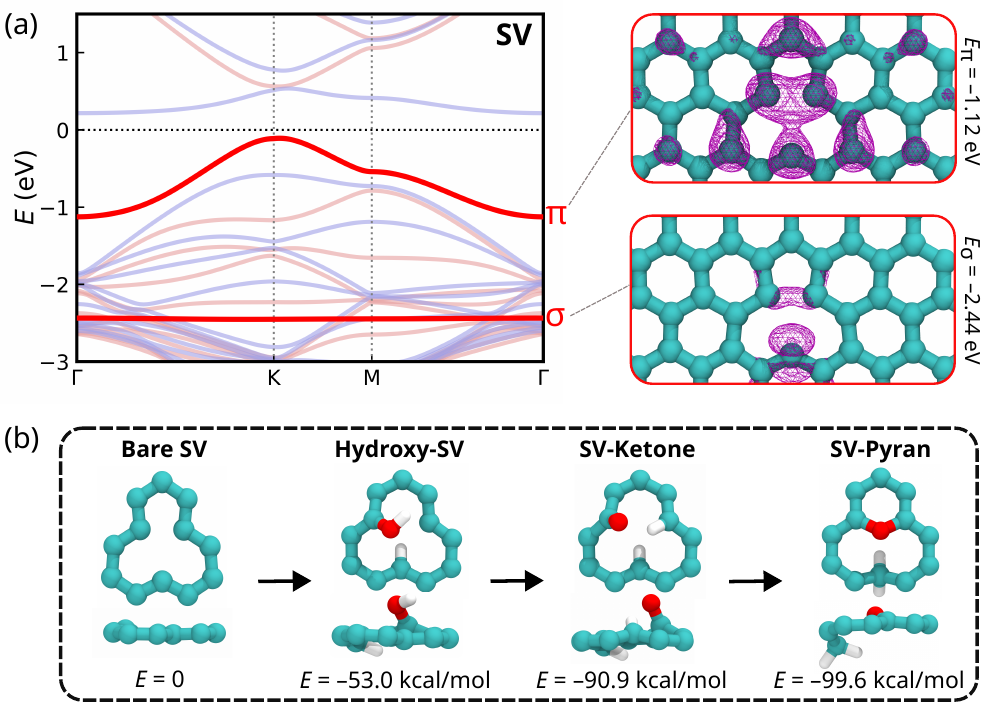}
\caption{Overview of the graphene single vacancy.
\textbf{(a)} Electronic structure of the graphene single vacancy. 
Plot shows the near-Fermi level band structure of the single vacancy calculated from 7x7 hexagonal cells.
The diradical $\sigma$ and $\pi$ bands are highlighted and shown alongside the corresponding molecular orbitals.
\textbf{(b)} Products of water dissociation at the graphene single vacancy.
Structures are shown alongside their respective potential energies (see Figure S4 of the Supporting Information).
}
\label{fig:sv_overview}
\end{figure*}

To address these questions, we begin by characterizing graphene surfaces, which are widely used in membranes, electrodes, and nanofluidic devices in permanent contact with liquid water.
Under pristine conditions, graphene does not dissociate water due to weak graphene-water interactions \cite{Brandenburg2019}.
Under realistic operating conditions, however, graphene is not a pristine sheet of regularly repeating carbon atoms;
it contains vacancies and edge sites that arise during synthesis, from chemical reactions, or from ionizing radiation \cite{Vinogradov2011,Krasheninnikov2010,Zhao2023}.
These defects have been observed experimentally \cite{Hashimoto2004,Meyer2008,Cancado2011,Mao2016,Zhang2016} and are known to alter the electronic, mechanical, and chemical properties of the graphene sheet \cite{Son2006,Banhart2011,Thiemann2025}. 
One of the most important of these defects is the graphene single vacancy (SV), a  diradical with two unpaired electrons (spin moment = $2 \: \mathrm{\mu_B}$) created by removing a single carbon atom at a cost of 7.9 eV \cite{Valencia2017,Formation2013}.
One of its unpaired electrons is located in a localized $\sigma$ orbital ($-2.4$~eV), while the other resides in an out-of-plane $\pi$ orbital ($-1.1$~eV)(see Figure \ref{fig:sv_overview}a) \cite{Yazyev2007,Palacios2012,Padmanabhan2016,Valencia2017}.
Together these defect states make the SV an effective reactive center, capable of both donating and accepting electron density from an incoming adsorbate -- a Janus-style defect. 
For the case of water dissociation, gas-phase density functional theory (DFT) calculations have confirmed that single water molecules chemisorb at the SV, proceeding through a concerted transition state to form a series of increasingly stable products \cite{Cabrera-Sanfelix2007,PhysRevLett.95.136105,Liang2021}.
This overall chemisorption process is shown in Figure \ref{fig:sv_overview}b. %
Water first attacks the bare vacancy and forms a Hydroxy-SV species containing C--OH and C--H groups.
The OH group then transfers a proton to the remaining carbon with a dangling bond to form a SV-Ketone species.
Finally, this ketone undergoes reorganization to form a SV-Pyran motif. 
Overall, the SV + \ch{H2O} reaction is highly exothermic, with the final products being roughly 100 kcal/mol more stable than the bare SV and physisorbed water.

The dissociation of a single water molecule at the graphene SV is understood \cite{Cabrera-Sanfelix2007,PhysRevLett.95.136105,Liang2021}. 
However, a description of this process under bulk solvation conditions is missing. 
This is notable given the capacity for surrounding water molecules to alter both the mechanism and energetics of surface water dissociation \cite{Michaelides2003,Michaelides2004,Tocci2014,CalegariFreeEnergy}. 
Traditional DFT approaches, which have been used extensively to characterize the gas-phase SV reactivity, are too costly for modeling fully solvated systems over the nanosecond simulation times required for enhanced sampling.
Classical force fields, meanwhile, cannot adequately describe bond-making and bond-breaking.

To overcome these limitations, we train a MACE interatomic potential on spin-polarized hybrid DFT data -- specifically, revPBE0-D3 -- to run enhanced-sampling simulations. 
Hybrid DFT is essential for this goal.
Semi-local functionals suffer more severely from self-interaction errors that cause delocalized $\pi$ bands to cross the Fermi level, artificially reducing the magnetization from its expected value of 2 $\mathrm{\mu_B}$ (see Figures S1 and S2 of the Supporting Information) \cite{Valencia2017}.
Combining our machine learning interatomic potential (MLIP) with multiple-walker well-tempered metadynamics, we enable multi-nanosecond trajectories of complex, reactive vacancy systems at a spin-polarized hybrid DFT level of theory (see Computational Methods for details).

\begin{figure*}[p]
\centering
    \includegraphics[width=0.85\textwidth]{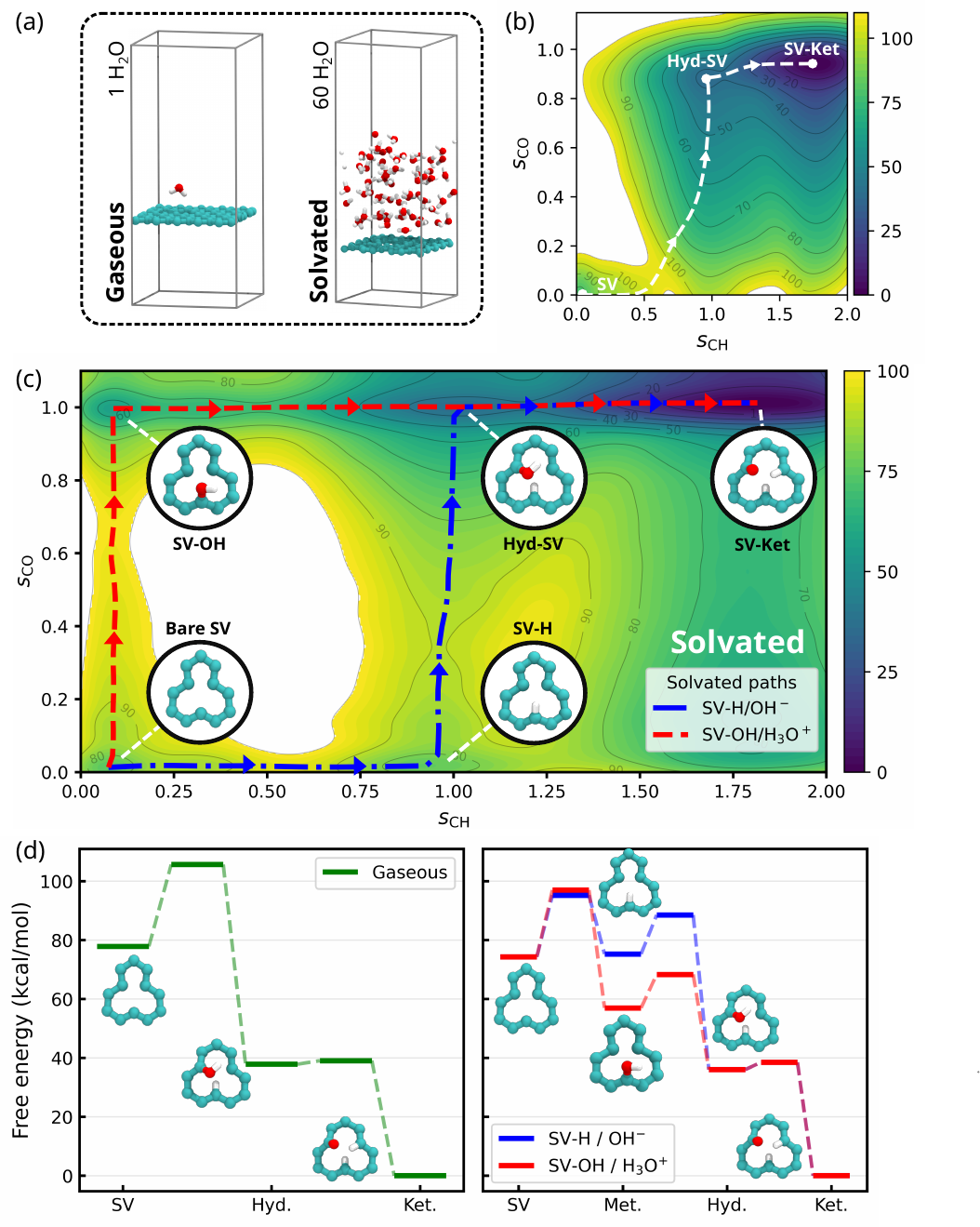}
\caption{Solvation modifies the reactivity of the graphene single vacancy.
\textbf{(a)} System setups for gaseous and solvated metadynamics simulations. 
\textbf{(b)} Free energy profile showing the SV-mediated decomposition of water under gaseous conditions (single water molecule). 
Free energies are shown as a function of C--H and C--O coordination numbers and plotted in kcal/mol.
The minimum energy pathway connecting reactant with product is shown by the white dashed line. 
\textbf{(c)} Free energy profile showing water decomposition under solvated conditions. 
Free energies are shown as a function of C--H and C--O, with the two viable reaction pathways shown in blue and red. 
Representative snapshots for these states are shown on the figure. 
\textbf{(d)} Free energy state diagram connecting the single vacancy to the SV-Ketone state for both gaseous (left) and solvated (right) conditions. 
}
\label{fig:sv_react}
\end{figure*}

We characterize the SV + \ch{H2O} reaction by computing 2D free energy surfaces as a function of the collective variables $s_\mathrm{CH}$ and $s_\mathrm{CO}$, which track the coordination of the SV site to hydrogen and oxygen atoms, respectively (see Computational Methods for more details).
These free energy surfaces are shown in Figures \ref{fig:sv_react}b and \ref{fig:sv_react}c, with the relevant minima and transition states shown as state-energy profiles in Figure \ref{fig:sv_react}d.
For this analysis, we show the dissociation process only up to the SV-Ketone state: this species is kinetically stabilized under ambient conditions, separated from the thermodynamically more stable SV-Pyran by a forward free energy barrier of $\sim 45$~kcal/mol.
This final SV-Ketone $\rightarrow$ SV-Pyran transformation is therefore treated separately in Figure S5 of the Supporting Information.

The gaseous reaction presents a relatively simple free energy landscape, with a single reaction channel connecting the bare SV to the chemisorbed product (Figure~\ref{fig:sv_react}b/d). 
Water attacks the vacancy over a 28.4 kcal/mol barrier in a concerted step that simultaneously breaks an O–H bond and forms C–OH and C–H bonds at the defect site, yielding the Hydroxy-SV intermediate. 
This step is rate-limiting for the dissociation process, and the Hydroxy-SV easily decomposes to form the SV-Ketone over a very small barrier. 
Overall, gaseous water dissociation is strongly exergonic, with the SV-Ketone lying 78 kcal/mol below the bare SV + \ch{H2O} reactant. 
This value is $\sim 13$ kcal/mol smaller in magnitude than the potential energy difference ($\Delta E \sim 91$ kcal/mol, see Figure~S4), a difference that we can attribute to the entropic cost of immobilising a free molecule at the surface \cite{Campbell2012}.

The solvated free energy landscape is more intricate (see Figure \ref{fig:sv_react}c/d). 
Two distinct reaction channels now connect the bare SV to the chemisorbed products, corresponding to two modes of water attack on the vacancy: 
\begin{equation}
    \mathrm{SV \: + \: H_2O \: \rightarrow SV\text{-}H + OH^-}_\text{(aq)}
\end{equation}
\begin{equation}
    \mathrm{SV \: + \: 2H_2O \: \rightarrow SV\text{-}OH + H_3O^+}_\text{(aq)}
\end{equation}
In the first pathway, the SV abstracts a proton from the nearby water to form an SV--H intermediate and an adjacent solvated hydroxide ion. 
In the second, water attacks the vacancy oxygen-first to form SV--OH, transferring one of its protons to the surrounding media via the Grotthus mechanism. 
Both processes occur with comparable forward free energy barriers -- indicative of an early transition state governed primarily by O--H bond cleavage -- but with the SV--OH/\ch{H3O+} intermediate being $\sim 20$ kcal/mol more stable in free energy. 
These forward barriers (21.9 kcal/mol for SV--H/\ch{OH-} formation, and 23.7 kcal/mol for SV--OH/\ch{H3O+}) are overall lower in free energy than for the concerted gas-phase process (28.4 kcal/mol).
From either intermediate, the system proceeds to the Hydroxy-SV and then to the SV-Ketone, which lies 74 kcal/mol below the solvated bare SV.

\begin{figure*}[t]
\centering
    \includegraphics[width=0.70\textwidth]{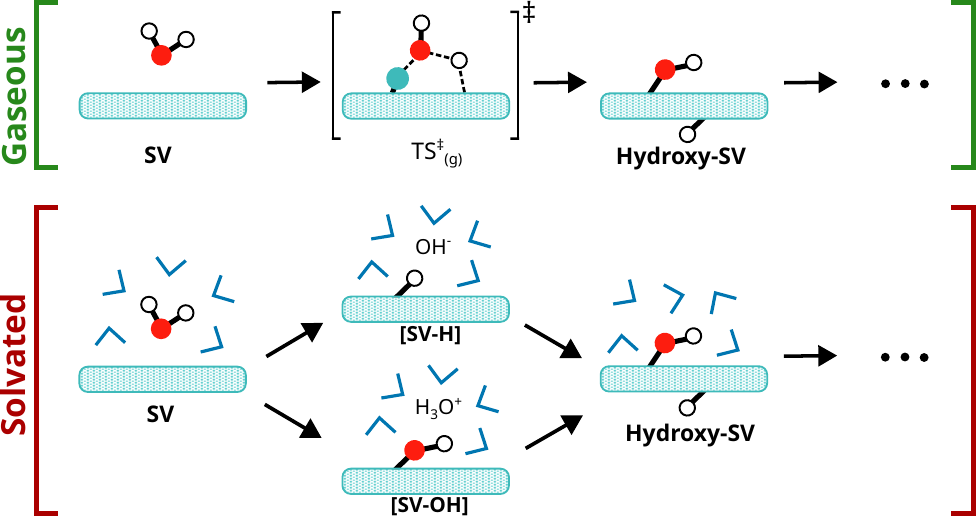}
\caption{Different scenarios for water decomposition at a solid surface. Top (gaseous mechanism): Single step dissociative adsorption. Bottom (solvated mechanisms):  Dissociative adsorption can proceed either via a chemisorbed OH and a solvated \ch{H3O+} (acidic route) or a chemisorbed H and a solvated \ch{OH-} (basic route). Through these two distinct mechanisms new, lower energy channels for decomposition are available. 
}
\label{fig:sv_mechanism}
\end{figure*}

The contrast between gaseous and solvated water dissociation arises directly from the surrounding liquid. 
Solvent molecules stabilize the charged \ch{H3O+} and \ch{OH-} fragments through hydrogen bonding and dielectric screening, opening up heterolytic reaction channels that have no gas-phase counterpart. 
Under gaseous conditions, where no charge stabilization is possible, dissociation proceeds through a concerted transition state that forces a hydrogen atom through the graphene plane.
The overall reaction mechanisms for gaseous and solvated water dissociation are shown in Figure \ref{fig:sv_mechanism}.
The dual reactivity of the solvated Janus-like SV, accepting either the hydrogen or the hydroxyl fragment of the dissociating water molecule, is reminiscent of the ambiphilic character of carbenes in molecular chemistry \cite{Bourissou1999}.
The graphene SV possesses a similar pair of $\sigma$ and $\pi$ orbitals, and in this sense acts as a surface analogue of molecular carbenes.
The greater energetic stability of the SV-OH intermediate over SV-H suggests that the acidic pathway is thermodynamically favored, which may produce a local excess of hydronium ions at the defect sites. 
This chemical contribution to interfacial acidity would add to the proton accumulation already observed at pristine graphene surfaces \cite{Advincula2025}.

\begin{figure*}[t!]
\centering
    \includegraphics[width=0.95\textwidth]{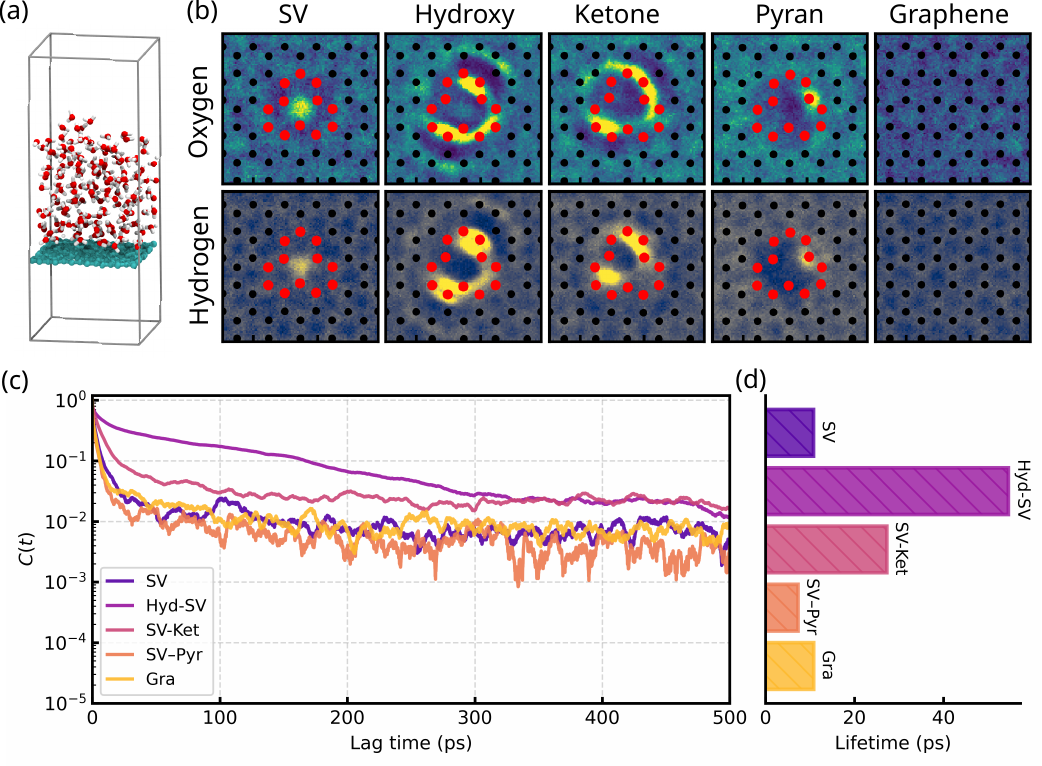}
\caption{Defect chemistry controls water residence lifetimes.
\textbf{(a)} Extended defective graphene system used for characterizing water adsorption at the vacancy site. 
\textbf{(b)} 2D adsorption profiles showing the distribution of oxygen (top panels) and hydrogen (bottom panels) within the water contact layer.
Average carbon positions are shown as black dots, with the atoms outlining the vacancy site highlighted in red.
\textbf{(c)} Plot of the correlation function $c(t)$ as a function of lag time for each defect type. 
\textbf{(d)} Lifetime estimates obtained from integrating under the $c(t)$ curve for each vacancy type.
}
\label{fig:nvt}
\end{figure*}

The competing chemisorption pathways highlight the important contribution of solvent molecules to surface water dissociation. 
The result is a series of structurally distinct intermediates, each presenting different functional groups to the solvent that modify the local water environment in different ways. 
We investigate this using 2D adsorption profiles and water residence lifetimes, which are shown in Figure \ref{fig:nvt}.
Hydrogen bonding between the chemisorbed groups and the surrounding liquid is the primary impetus of enhanced water adsorption, consistent with previous \textit{ab initio} molecular dynamics studies of defective graphene-water interfaces \cite{Tocci2016}.
The effect is strongly product-dependent. 
The Hydroxy-SV ($\tau$~=~55~ps) and SV-Ketone ($\tau$~=~27~ps) states, both of which present hydrogen-bond-active O–H or C=O groups to the liquid, display water residence times several times greater than either the bare SV or the SV-Pyran, which are comparable to the pristine graphene lifetime.
Chemisorption therefore not only transforms the local electronic structure of the defect, but also converts it from a hydrophobic to a hydrophilic site while simultaneously passivating the SV against further chemical attack from water. 
This suggests that controlled defect chemistry could offer a route to modulating wettability and friction at graphene--water interfaces, and that quantitative modeling of these phenomena requires moving beyond the pristine graphene approximation.

\vspace{5mm}

\noindent In conclusion, we show that solvation qualitatively transforms  the mechanism of water dissociation at defects.
Whereas the gas-phase reaction proceeds via a single concerted channel, solvated dissociation is a bimodal, stepwise process in which the surrounding liquid stabilizes charged intermediates and opens heterolytic pathways that are absent in vacuum. 
These solvated reaction routes offer lower forward barriers and yield chemisorbed products that, in turn, enhance local water adsorption at the defect site.

Our work highlights the importance of explicit solvation in simulating reactive interfacial chemical systems. 
This has been established for metals and metal oxides, but our results demonstrate that it extends to 2D carbon surfaces, where a single atomic vacancy is sufficient to transform the local environment from physisorption to reactive chemisorption. 
The two Janus-like acidic and basic pathways identified here, as well as the thermodynamic preference for the acidic channel, suggest that solvated graphene vacancies may act as short-lived, local sources of hydronium, contributing to interfacial acidity beyond what is observed at pristine graphene surfaces \cite{Advincula2025}.
More broadly, direct comparisons between gas-phase and solvated water dissociation remain rare for any surface. 
The qualitative changes we observe, i.e., solvation opening competing ionic channels, stabilizing intermediates, and lowering barriers, are unlikely to be unique to carbon and may require reassessment of reaction mechanisms previously characterized only under vacuum conditions.
We expect that these effects will also be relevant to other 2D material interfaces, for example, for monolayer hBN which exhibits strong surface charging effects \cite{WanghBNJACS}.

Looking forward, the distinct states identified here, each with different water adsorption properties and residence times, suggest that controlled defect engineering could offer a route to tuning wettability, friction, and selective ion transport in 2D-membrane materials. 
An important next step is to understand cooperative effects and whether proximal vacancies react independently or exhibit coupled reactivity that could be exploited for selective surface chemistry \cite{Thiemann2025}. 
More generally, our work demonstrates that machine-learned potentials trained on hybrid DFT, combined with enhanced-sampling techniques, now make it feasible to resolve the full free energy landscape of rare reactive events at explicitly solvated interfaces.
This approach is readily transferable to other defective 2D materials and reactive solid--liquid systems.

\subsection*{Computational Methods}
In this work, we develop an equivariant MACE potential at a hybrid DFT level of theory.
Training structures were generated using exhaustive sampling through the use of foundation MACE models (MP0)\cite{Batatia2023}, through AIMD, and from preliminary NEB and umbrella integration runs. 
In total, around 9000 structures were generated, with configurations including defects under solvated, partially solvated, and gaseous conditions. 
For each configuration, single-point DFT calculations were performed using FHI-aims \cite{Blum2009,Ren2012,Lehtola2018}. 
Spin-unrestricted Kohn-Sham calculations were used for configurations containing undercoordinated carbon atoms; spin-restricted calculations were used otherwise.
In this way, we selected the lowest-energy electronic states such that we were sampling the adiabatic energy surface. 
For SV and other triplet-state calculations, initial magnetizations of $0.5-1.0$ $\mu_\mathrm{B}$ were applied to the three nearest carbon atoms surrounding the vacancy site. 
Hybrid calculations were performed using the revPBE0 functional augmented by Grimme's D3 corrections (zero damping) \cite{PhysRevLett.77.3865,PhysRevLett.80.890,doi:10.1063/1.478522,doi:10.1063/1.3382344}. 
For these calculations, we employed the intermediate orbital basis with a $4 \times 4 \times 1$ k-grid.

The MACE potential was generated via a two-step procedure. 
First, a base model was generated at a GGA level of theory (revPBE-D3\cite{PhysRevLett.77.3865,PhysRevLett.80.890,doi:10.1063/1.3382344}, $\sim$ 7000 labeled structures).
This model was then transfer-learned to the revPBE0-D3 level via transfer learning on a smaller subset of labeled structures ($\sim$ 2000).
This final hybrid-level MACE model was generated with 128 channels and a maximal message equivariance of $L=1$.
A radial cutoff of 5 \AA{} was selected, which equates to an effective receptive field of 10 \AA{} after message passing. 
The final model exhibited training RMSEs of 1.3 meV/atom and 30.6 meV/\AA{} for the energies and forces, respectively.
Model performance was evaluated through force error analysis (see Figure S3) and through NEB calculations across the water dissociation pathway, where MACE energies and barriers agree with those of revPBE0-D3 reference values to within $2-3$~kcal/mol (see Figure S4).

Well-tempered metadynamics was performed by coupling \texttt{LAMMPS} with the multiple-walker setup available in \texttt{PLUMED} \cite{Thompson2022,Barducci2008,Raiteri2006,PLUMED,PLUMED2}.
For the gas-phase reaction, 4 walkers were initialized from various states spanning the reaction coordinate using a 12.350 \AA{} $\times$ 12.834 \AA{} $\times$ 30.000 \AA{} simulation cell containing a single  water molecule (see Figure \ref{fig:sv_react}a). 
For the solvated reaction, 8 walkers were similarly initialized using a 12.350 \AA{} $\times$ 12.834 \AA{} $\times$ 36.000 \AA{} cell containing $\sim 60$ water molecules (see Figure \ref{fig:sv_react}a). 
Simulations were performed using the \textit{NVT} ensemble (Nosé-Hoover-style) at 300 K and with a time constant of 100 fs. 
The individual time step was set to 0.5 fs. 
We chose two collective variables to probe this reaction: a C-H coordination number that tracks the number of hydrogens bound at the single vacancy; and a C-O coordination number that tracks the number of oxygen atoms bound at the site. 
Restraints were appropriately applied to restrict reaction to a single water molecule at any one time reacting directly at the single vacancy. 
Some side reactions were observed during the course of these simulations (e.g., oxygen slipping onto an adjacent, non-defect carbon) but these did not perturb the defects state energies and minima under consideration. 
Convergence of the metadynamics runs was monitored by tracking the relative free energies of the various chemisorbed states. 
In total, the gaseous and solvated metadynamics simulations were run for a cumulative time of 6 ns and 40 ns, respectively. 
Unbiased $NVT$ simulations, used to characterize the physisorption of water at the graphene defects, were performed over 2 ns using extended simulation cells (17.290 \AA{} $\times$ 17.112 \AA{} $\times$ 45.00 \AA{}) consisting of $\sim 160$ water molecules (see Figure \ref{fig:nvt}a). 
The following correlation function $c(t)$ was used to analyze water adsorption and residence times: $c(t) = \langle h(0)h(t) \rangle / \langle h \rangle$ where $h$ is 1 for a water adsorbed at the defect site (within 3.5 \AA{} of the defect atoms) and 0 otherwise.

\section*{Acknowledgements}
\noindent SGHB is supported by the Syntech CDT and funded by EPSRC (Grant No.\ EP/S024220/1).
ICP acknowledges the financial support from the Gates Cambridge Trust (Grant No.\ OPP1144).
FB acknowledges support from the Alexander von Humboldt Foundation through a Feodor Lynen Research Fellowship, from the Isaac Newton Trust through an Early Career Fellowship, and from Churchill College, Cambridge, through a Postdoctoral By-Fellowship. 
AM acknowledges support from the European Union under the “n-AQUA” European Research Council project (Grant No. 101071937).
CS also acknowledges support from the Isaac Newton Trust G122390 and the UKRI reference EP/V062654/1.
We are grateful for computational support and resources from the UK Materials and Molecular Modeling Hub which is partially funded by EPSRC (Grant Nos. EP/P020194/1 and EP/T022213/1).
We are also grateful for computational support and resources from the UK national high-performance computing service, Advanced Research Computing High End Resource (ARCHER2) and the Swiss National Supercomputing Centre under project s1209.
Access for both the UK Materials and Molecular Modeling Hub and ARCHER2 were obtained via the UK Car-Parrinello consortium, funded by EPSRC grant reference EP/P022561/1.
Access to CSD3 was obtained through a University of Cambridge EPSRC Core Equipment Award (EP/X034712/1).
We also acknowledge the EuroHPC Joint Undertaking for awarding project ID EHPC-REG-2024R02-130 access to Leonardo at CINECA (Italy), and project ID EHPC-REG-2025R02-112 access to JUPITER at Jülich (Germany).

\section*{Supporting Information}
\noindent Supporting Information Available: DFT calculations, model development, validation protocols, and free energy profiles for the SV-Pyran formation.

\section*{Data Availability}
\noindent  Models, figure data, and input files will be made openly available on GitHub upon acceptance of this manuscript.
Similarly, training structures and data will be made available on Zenodo.


\begin{mcitethebibliography}{65}
\providecommand*\natexlab[1]{#1}
\providecommand*\mciteSetBstSublistMode[1]{}
\providecommand*\mciteSetBstMaxWidthForm[2]{}
\providecommand*\mciteBstWouldAddEndPuncttrue
  {\def\EndOfBibitem{\unskip.}}
\providecommand*\mciteBstWouldAddEndPunctfalse
  {\let\EndOfBibitem\relax}
\providecommand*\mciteSetBstMidEndSepPunct[3]{}
\providecommand*\mciteSetBstSublistLabelBeginEnd[3]{}
\providecommand*\EndOfBibitem{}
\mciteSetBstSublistMode{f}
\mciteSetBstMaxWidthForm{subitem}{(\alph{mcitesubitemcount})}
\mciteSetBstSublistLabelBeginEnd
  {\mcitemaxwidthsubitemform\space}
  {\relax}
  {\relax}

\bibitem[Carrasco \latin{et~al.}(2012)Carrasco, Hodgson, and
  Michaelides]{Carrasco2012}
Carrasco,~J.; Hodgson,~A.; Michaelides,~A. A molecular perspective of water at
  metal interfaces. \emph{Nature Materials} \textbf{2012}, \emph{11},
  667--674\relax
\mciteBstWouldAddEndPuncttrue
\mciteSetBstMidEndSepPunct{\mcitedefaultmidpunct}
{\mcitedefaultendpunct}{\mcitedefaultseppunct}\relax
\EndOfBibitem
\bibitem[Bj\"orneholm \latin{et~al.}(2016)Bj\"orneholm, Hansen, Hodgson, Liu,
  Limmer, Michaelides, Pedevilla, Rossmeisl, Shen, Tocci, Tyrode, Walz, Werner,
  and Bluhm]{Olle2016}
Bj\"orneholm,~O.; Hansen,~M.~H.; Hodgson,~A.; Liu,~L.-M.; Limmer,~D.~T.;
  Michaelides,~A.; Pedevilla,~P.; Rossmeisl,~J.; Shen,~H.; Tocci,~G.
  \latin{et~al.}  Water at Interfaces. \emph{Chemical Reviews} \textbf{2016},
  \emph{116}, 7698--7726\relax
\mciteBstWouldAddEndPuncttrue
\mciteSetBstMidEndSepPunct{\mcitedefaultmidpunct}
{\mcitedefaultendpunct}{\mcitedefaultseppunct}\relax
\EndOfBibitem
\bibitem[Ruscic \latin{et~al.}(2002)Ruscic, Wagner, Harding, Asher, Feller,
  Dixon, Peterson, Song, Qian, Ng, Liu, Chen, and Schwenke]{Ruscic2002}
Ruscic,~B.; Wagner,~A.~F.; Harding,~L.~B.; Asher,~R.~L.; Feller,~D.;
  Dixon,~D.~A.; Peterson,~K.~A.; Song,~Y.; Qian,~X.; Ng,~C.-Y. \latin{et~al.}
  On the Enthalpy of Formation of Hydroxyl Radical and Gas-Phase Bond
  Dissociation Energies of Water and Hydroxyl. \emph{The Journal of Physical
  Chemistry A} \textbf{2002}, \emph{106}, 2727--2747\relax
\mciteBstWouldAddEndPuncttrue
\mciteSetBstMidEndSepPunct{\mcitedefaultmidpunct}
{\mcitedefaultendpunct}{\mcitedefaultseppunct}\relax
\EndOfBibitem
\bibitem[Phatak \latin{et~al.}(2009)Phatak, Delgass, Ribeiro, and
  Schneider]{Phatak2009}
Phatak,~A.~A.; Delgass,~W.~N.; Ribeiro,~F.~H.; Schneider,~W.~F. Density
  Functional Theory Comparison of Water Dissociation Steps on Cu, Au, Ni, Pd,
  and Pt. \emph{The Journal of Physical Chemistry C} \textbf{2009}, \emph{113},
  7269--7276\relax
\mciteBstWouldAddEndPuncttrue
\mciteSetBstMidEndSepPunct{\mcitedefaultmidpunct}
{\mcitedefaultendpunct}{\mcitedefaultseppunct}\relax
\EndOfBibitem
\bibitem[Subbaraman \latin{et~al.}(2011)Subbaraman, Tripkovic, Strmcnik, Chang,
  Uchimura, Paulikas, Stamenkovic, and Markovic]{Subbaraman2011}
Subbaraman,~R.; Tripkovic,~D.; Strmcnik,~D.; Chang,~K.-C.; Uchimura,~M.;
  Paulikas,~A.~P.; Stamenkovic,~V.; Markovic,~N.~M. Enhancing Hydrogen
  Evolution Activity in Water Splitting by Tailoring Li+-Ni(OH)2-Pt Interfaces.
  \emph{Science} \textbf{2011}, \emph{334}, 1256--1260\relax
\mciteBstWouldAddEndPuncttrue
\mciteSetBstMidEndSepPunct{\mcitedefaultmidpunct}
{\mcitedefaultendpunct}{\mcitedefaultseppunct}\relax
\EndOfBibitem
\bibitem[Rodriguez \latin{et~al.}(2007)Rodriguez, Ma, Liu, Hrbek, Evans, and
  Pérez]{Rodriguez2007}
Rodriguez,~J.~A.; Ma,~S.; Liu,~P.; Hrbek,~J.; Evans,~J.; Pérez,~M. Activity of
  CeOx and TiOx Nanoparticles Grown on Au(111) in the Water-Gas Shift Reaction.
  \emph{Science} \textbf{2007}, \emph{318}, 1757--1760\relax
\mciteBstWouldAddEndPuncttrue
\mciteSetBstMidEndSepPunct{\mcitedefaultmidpunct}
{\mcitedefaultendpunct}{\mcitedefaultseppunct}\relax
\EndOfBibitem
\bibitem[Fujishima and Honda(1972)Fujishima, and Honda]{Fujishima1972}
Fujishima,~A.; Honda,~K. Electrochemical Photolysis of Water at a Semiconductor
  Electrode. \emph{Nature} \textbf{1972}, \emph{238}, 37--38\relax
\mciteBstWouldAddEndPuncttrue
\mciteSetBstMidEndSepPunct{\mcitedefaultmidpunct}
{\mcitedefaultendpunct}{\mcitedefaultseppunct}\relax
\EndOfBibitem
\bibitem[Maeda \latin{et~al.}(2006)Maeda, Teramura, Lu, Takata, Saito, Inoue,
  and Domen]{Maeda2006}
Maeda,~K.; Teramura,~K.; Lu,~D.; Takata,~T.; Saito,~N.; Inoue,~Y.; Domen,~K.
  Photocatalyst releasing hydrogen from water. \emph{Nature} \textbf{2006},
  \emph{440}, 295\relax
\mciteBstWouldAddEndPuncttrue
\mciteSetBstMidEndSepPunct{\mcitedefaultmidpunct}
{\mcitedefaultendpunct}{\mcitedefaultseppunct}\relax
\EndOfBibitem
\bibitem[Borup \latin{et~al.}(2007)Borup, Meyers, Pivovar, Kim, Mukundan,
  Garland, Myers, Wilson, Garzon, Wood, Zelenay, More, Stroh, Zawodzinski,
  Boncella, McGrath, Inaba, Miyatake, Hori, Ota, Ogumi, Miyata, Nishikata,
  Siroma, Uchimoto, Yasuda, ichi Kimijima, and Iwashita]{Borup2007}
Borup,~R.; Meyers,~J.; Pivovar,~B.; Kim,~Y.~S.; Mukundan,~R.; Garland,~N.;
  Myers,~D.; Wilson,~M.; Garzon,~F.; Wood,~D. \latin{et~al.}  Scientific
  Aspects of Polymer Electrolyte Fuel Cell Durability and Degradation.
  \emph{Chemical Reviews} \textbf{2007}, \emph{107}, 3904--3951\relax
\mciteBstWouldAddEndPuncttrue
\mciteSetBstMidEndSepPunct{\mcitedefaultmidpunct}
{\mcitedefaultendpunct}{\mcitedefaultseppunct}\relax
\EndOfBibitem
\bibitem[Henderson(2002)]{Henderson2002}
Henderson,~M.~A. The interaction of water with solid surfaces: fundamental
  aspects revisited. \emph{Surface Science Reports} \textbf{2002}, \emph{46},
  1--308\relax
\mciteBstWouldAddEndPuncttrue
\mciteSetBstMidEndSepPunct{\mcitedefaultmidpunct}
{\mcitedefaultendpunct}{\mcitedefaultseppunct}\relax
\EndOfBibitem
\bibitem[Michaelides \latin{et~al.}(2003)Michaelides, Alavi, and
  King]{Michaelides2003}
Michaelides,~A.; Alavi,~A.; King,~D.~A. Different Surface Chemistries of Water
  on Ru\{0001\}: From Monomer Adsorption to Partially Dissociated Bilayers.
  \emph{Journal of the American Chemical Society} \textbf{2003}, \emph{125},
  2746--2755\relax
\mciteBstWouldAddEndPuncttrue
\mciteSetBstMidEndSepPunct{\mcitedefaultmidpunct}
{\mcitedefaultendpunct}{\mcitedefaultseppunct}\relax
\EndOfBibitem
\bibitem[Michaelides \latin{et~al.}(2004)Michaelides, Alavi, and
  King]{Michaelides2004}
Michaelides,~A.; Alavi,~A.; King,~D.~A. Insight into H2O-ice adsorption and
  dissociation on metal surfaces from first-principles simulations.
  \emph{Physical Review B} \textbf{2004}, \emph{69}, 113404\relax
\mciteBstWouldAddEndPuncttrue
\mciteSetBstMidEndSepPunct{\mcitedefaultmidpunct}
{\mcitedefaultendpunct}{\mcitedefaultseppunct}\relax
\EndOfBibitem
\bibitem[Desai and Neurock(2003)Desai, and Neurock]{Desai2003}
Desai,~S.~K.; Neurock,~M. First-principles study of the role of solvent in the
  dissociation of water over a Pt-Ru alloy. \emph{Physical Review B}
  \textbf{2003}, \emph{68}, 75420\relax
\mciteBstWouldAddEndPuncttrue
\mciteSetBstMidEndSepPunct{\mcitedefaultmidpunct}
{\mcitedefaultendpunct}{\mcitedefaultseppunct}\relax
\EndOfBibitem
\bibitem[Donadio \latin{et~al.}(2012)Donadio, Ghiringhelli, and
  Site]{Donadio2012}
Donadio,~D.; Ghiringhelli,~L.~M.; Site,~L.~D. Autocatalytic and Cooperatively
  Stabilized Dissociation of Water on a Stepped Platinum Surface. \emph{Journal
  of the American Chemical Society} \textbf{2012}, \emph{134},
  19217--19222\relax
\mciteBstWouldAddEndPuncttrue
\mciteSetBstMidEndSepPunct{\mcitedefaultmidpunct}
{\mcitedefaultendpunct}{\mcitedefaultseppunct}\relax
\EndOfBibitem
\bibitem[Pek\"oz and Donadio(2017)Pek\"oz, and Donadio]{Donadio2017}
Pek\"oz,~R.; Donadio,~D. Dissociative Adsorption of Water at (211) Stepped
  Metallic Surfaces by First-Principles Simulations. \emph{The Journal of
  Physical Chemistry C} \textbf{2017}, \emph{121}, 16783--16791\relax
\mciteBstWouldAddEndPuncttrue
\mciteSetBstMidEndSepPunct{\mcitedefaultmidpunct}
{\mcitedefaultendpunct}{\mcitedefaultseppunct}\relax
\EndOfBibitem
\bibitem[Schaub \latin{et~al.}(2001)Schaub, Thostrup, Lopez, Lægsgaard,
  Stensgaard, Nørskov, and Besenbacher]{Schaub2001}
Schaub,~R.; Thostrup,~P.; Lopez,~N.; Lægsgaard,~E.; Stensgaard,~I.;
  Nørskov,~J.~K.; Besenbacher,~F. Oxygen Vacancies as Active Sites for Water
  Dissociation on Rutile TiO2(110). \emph{Physical Review Letters}
  \textbf{2001}, \emph{87}, 266104\relax
\mciteBstWouldAddEndPuncttrue
\mciteSetBstMidEndSepPunct{\mcitedefaultmidpunct}
{\mcitedefaultendpunct}{\mcitedefaultseppunct}\relax
\EndOfBibitem
\bibitem[Bikondoa \latin{et~al.}(2006)Bikondoa, Pang, Ithnin, Muryn, Onishi,
  and Thornton]{Bikondoa2006}
Bikondoa,~O.; Pang,~C.~L.; Ithnin,~R.; Muryn,~C.~A.; Onishi,~H.; Thornton,~G.
  Direct visualization of defect-mediated dissociation of water on TiO2(110).
  \emph{Nature Materials} \textbf{2006}, \emph{5}, 189--192\relax
\mciteBstWouldAddEndPuncttrue
\mciteSetBstMidEndSepPunct{\mcitedefaultmidpunct}
{\mcitedefaultendpunct}{\mcitedefaultseppunct}\relax
\EndOfBibitem
\bibitem[Andrade \latin{et~al.}(2020)Andrade, Ko, Zhang, Car, and
  Selloni]{CalegariFreeEnergy}
Andrade,~M. F.~C.; Ko,~H.-Y.; Zhang,~L.; Car,~R.; Selloni,~A. Free energy of
  proton transfer at the water–TiO2 interface from ab initio deep potential
  molecular dynamics. \emph{Chemical Science} \textbf{2020}, \emph{11},
  2335--2341\relax
\mciteBstWouldAddEndPuncttrue
\mciteSetBstMidEndSepPunct{\mcitedefaultmidpunct}
{\mcitedefaultendpunct}{\mcitedefaultseppunct}\relax
\EndOfBibitem
\bibitem[Zhuang \latin{et~al.}(2022)Zhuang, Bi, and Cheng]{Zhuang2022}
Zhuang,~Y.-B.; Bi,~R.-H.; Cheng,~J. Resolving the odd–even oscillation of
  water dissociation at rutile TiO2(110)–water interface by machine learning
  accelerated molecular dynamics. \emph{The Journal of Chemical Physics}
  \textbf{2022}, \emph{157}, 164701\relax
\mciteBstWouldAddEndPuncttrue
\mciteSetBstMidEndSepPunct{\mcitedefaultmidpunct}
{\mcitedefaultendpunct}{\mcitedefaultseppunct}\relax
\EndOfBibitem
\bibitem[Wen \latin{et~al.}(2023)Wen, Andrade, Liu, and Selloni]{WenTiO2PNAS}
Wen,~B.; Andrade,~M. F.~C.; Liu,~L.-M.; Selloni,~A. Water dissociation at the
  water–rutile TiO2(110) interface from ab initio-based deep neural network
  simulations. \emph{Proceedings of the National Academy of Sciences}
  \textbf{2023}, \emph{120}, e2212250120\relax
\mciteBstWouldAddEndPuncttrue
\mciteSetBstMidEndSepPunct{\mcitedefaultmidpunct}
{\mcitedefaultendpunct}{\mcitedefaultseppunct}\relax
\EndOfBibitem
\bibitem[Zeng \latin{et~al.}(2023)Zeng, Wodaczek, Liu, Stein, Hutter, Chen, and
  Cheng]{ZengTiO2NatComm}
Zeng,~Z.; Wodaczek,~F.; Liu,~K.; Stein,~F.; Hutter,~J.; Chen,~J.; Cheng,~B.
  Mechanistic insight on water dissociation on pristine low-index TiO2 surfaces
  from machine learning molecular dynamics simulations. \emph{Nature
  Communications} \textbf{2023}, \emph{14}, 6131\relax
\mciteBstWouldAddEndPuncttrue
\mciteSetBstMidEndSepPunct{\mcitedefaultmidpunct}
{\mcitedefaultendpunct}{\mcitedefaultseppunct}\relax
\EndOfBibitem
\bibitem[Tocci and Michaelides(2014)Tocci, and Michaelides]{Tocci2014}
Tocci,~G.; Michaelides,~A. Solvent-Induced Proton Hopping at a Water–Oxide
  Interface. \emph{The Journal of Physical Chemistry Letters} \textbf{2014},
  \emph{5}, 474--480\relax
\mciteBstWouldAddEndPuncttrue
\mciteSetBstMidEndSepPunct{\mcitedefaultmidpunct}
{\mcitedefaultendpunct}{\mcitedefaultseppunct}\relax
\EndOfBibitem
\bibitem[Zhang \latin{et~al.}(2016)Zhang, Wang, Liu, Liu, Dong, Zhuang, Chen,
  and Feng]{ZhangMoS2EES}
Zhang,~J.; Wang,~T.; Liu,~P.; Liu,~S.; Dong,~R.; Zhuang,~X.; Chen,~M.; Feng,~X.
  Engineering water dissociation sites in MoS2 nanosheets for accelerated
  electrocatalytic hydrogen production. \emph{Energy \& Environmental Science}
  \textbf{2016}, \emph{9}, 2789--2793\relax
\mciteBstWouldAddEndPuncttrue
\mciteSetBstMidEndSepPunct{\mcitedefaultmidpunct}
{\mcitedefaultendpunct}{\mcitedefaultseppunct}\relax
\EndOfBibitem
\bibitem[Wang \latin{et~al.}(2025)Wang, Luo, Advincula, Zhao, Esfandiar, Wu,
  Fong, Gao, Hazrah, Taniguchi, Schran, Nagata, Bocquet, Bocquet, Jiang,
  Michaelides, and Bonn]{WanghBNJACS}
Wang,~Y.; Luo,~H.; Advincula,~X.~R.; Zhao,~Z.; Esfandiar,~A.; Wu,~D.;
  Fong,~K.~D.; Gao,~L.; Hazrah,~A.~S.; Taniguchi,~T. \latin{et~al.}
  Spontaneous Surface Charging and Janus Nature of the Hexagonal Boron
  Nitride–Water Interface. \emph{Journal of the American Chemical Society}
  \textbf{2025}, \emph{147}, 30107--30116\relax
\mciteBstWouldAddEndPuncttrue
\mciteSetBstMidEndSepPunct{\mcitedefaultmidpunct}
{\mcitedefaultendpunct}{\mcitedefaultseppunct}\relax
\EndOfBibitem
\bibitem[Comtet \latin{et~al.}(2020)Comtet, Grosjean, Glushkov, Avsar,
  Watanabe, Taniguchi, Vuilleumier, Bocquet, and Radenovic]{ComtethBN}
Comtet,~J.; Grosjean,~B.; Glushkov,~E.; Avsar,~A.; Watanabe,~K.; Taniguchi,~T.;
  Vuilleumier,~R.; Bocquet,~M.-L.; Radenovic,~A. Direct observation of
  water-mediated single-proton transport between hBN surface defects.
  \emph{Nature Nanotechnology} \textbf{2020}, \emph{15}, 598--604\relax
\mciteBstWouldAddEndPuncttrue
\mciteSetBstMidEndSepPunct{\mcitedefaultmidpunct}
{\mcitedefaultendpunct}{\mcitedefaultseppunct}\relax
\EndOfBibitem
\bibitem[Joly \latin{et~al.}(2016)Joly, Tocci, Merabia, and
  Michaelides]{Tocci2016}
Joly,~L.; Tocci,~G.; Merabia,~S.; Michaelides,~A. Strong Coupling between
  Nanofluidic Transport and Interfacial Chemistry: How Defect Reactivity
  Controls Liquid–Solid Friction through Hydrogen Bonding. \emph{The Journal
  of Physical Chemistry Letters} \textbf{2016}, \emph{7}, 1381--1386\relax
\mciteBstWouldAddEndPuncttrue
\mciteSetBstMidEndSepPunct{\mcitedefaultmidpunct}
{\mcitedefaultendpunct}{\mcitedefaultseppunct}\relax
\EndOfBibitem
\bibitem[Scalfi \latin{et~al.}(2023)Scalfi, Becker, Netz, and
  Bocquet]{Scalfi2023}
Scalfi,~L.; Becker,~M.~R.; Netz,~R.~R.; Bocquet,~M.-L. Enhanced interfacial
  water dissociation on a hydrated iron porphyrin single-atom catalyst in
  graphene. \emph{Communications Chemistry} \textbf{2023}, \emph{6}, 236\relax
\mciteBstWouldAddEndPuncttrue
\mciteSetBstMidEndSepPunct{\mcitedefaultmidpunct}
{\mcitedefaultendpunct}{\mcitedefaultseppunct}\relax
\EndOfBibitem
\bibitem[Zhang \latin{et~al.}(2026)Zhang, Tian, Huang, Luo, and
  Chen]{Zhang2026}
Zhang,~X.; Tian,~J.; Huang,~Y.; Luo,~X.; Chen,~H. Taming Gas-Phase Radical
  Contributions to the Oxidative Dehydrogenation of Propane by Tailoring Defect
  Structures in Boron Nitride. \emph{ACS Catalysis} \textbf{2026}, \emph{16},
  16947--16957\relax
\mciteBstWouldAddEndPuncttrue
\mciteSetBstMidEndSepPunct{\mcitedefaultmidpunct}
{\mcitedefaultendpunct}{\mcitedefaultseppunct}\relax
\EndOfBibitem
\bibitem[Brandenburg \latin{et~al.}(2019)Brandenburg, Zen, Fitzner, Ramberger,
  Kresse, Tsatsoulis, Gr\"uneis, Michaelides, and Alf\`e]{Brandenburg2019}
Brandenburg,~J.~G.; Zen,~A.; Fitzner,~M.; Ramberger,~B.; Kresse,~G.;
  Tsatsoulis,~T.; Gr\"uneis,~A.; Michaelides,~A.; Alf\`e,~D. Physisorption of
  Water on Graphene: Subchemical Accuracy from Many-Body Electronic Structure
  Methods. \emph{The Journal of Physical Chemistry Letters} \textbf{2019},
  \emph{10}, 358--368\relax
\mciteBstWouldAddEndPuncttrue
\mciteSetBstMidEndSepPunct{\mcitedefaultmidpunct}
{\mcitedefaultendpunct}{\mcitedefaultseppunct}\relax
\EndOfBibitem
\bibitem[Vinogradov \latin{et~al.}(2011)Vinogradov, Schulte, Ng, Mikkelsen,
  Lundgren, Mårtensson, and Preobrajenski]{Vinogradov2011}
Vinogradov,~N.~A.; Schulte,~K.; Ng,~M.~L.; Mikkelsen,~A.; Lundgren,~E.;
  Mårtensson,~N.; Preobrajenski,~A.~B. Impact of Atomic Oxygen on the
  Structure of Graphene Formed on Ir(111) and Pt(111). \emph{The Journal of
  Physical Chemistry C} \textbf{2011}, \emph{115}, 9568--9577\relax
\mciteBstWouldAddEndPuncttrue
\mciteSetBstMidEndSepPunct{\mcitedefaultmidpunct}
{\mcitedefaultendpunct}{\mcitedefaultseppunct}\relax
\EndOfBibitem
\bibitem[Krasheninnikov and Nordlund(2010)Krasheninnikov, and
  Nordlund]{Krasheninnikov2010}
Krasheninnikov,~A.~V.; Nordlund,~K. Ion and electron irradiation-induced
  effects in nanostructured materials. \emph{Journal of Applied Physics}
  \textbf{2010}, \emph{107}, 71301\relax
\mciteBstWouldAddEndPuncttrue
\mciteSetBstMidEndSepPunct{\mcitedefaultmidpunct}
{\mcitedefaultendpunct}{\mcitedefaultseppunct}\relax
\EndOfBibitem
\bibitem[Zhao \latin{et~al.}(2023)Zhao, Chen, Zhang, Yi, Chen, Su, Niu, Zhang,
  and Long]{Zhao2023}
Zhao,~Z.; Chen,~H.; Zhang,~W.; Yi,~S.; Chen,~H.; Su,~Z.; Niu,~B.; Zhang,~Y.;
  Long,~D. Defect engineering in carbon materials for electrochemical energy
  storage and catalytic conversion. \emph{Materials Advances} \textbf{2023},
  \emph{4}, 835--867\relax
\mciteBstWouldAddEndPuncttrue
\mciteSetBstMidEndSepPunct{\mcitedefaultmidpunct}
{\mcitedefaultendpunct}{\mcitedefaultseppunct}\relax
\EndOfBibitem
\bibitem[Hashimoto \latin{et~al.}(2004)Hashimoto, Suenaga, Gloter, Urita, and
  Iijima]{Hashimoto2004}
Hashimoto,~A.; Suenaga,~K.; Gloter,~A.; Urita,~K.; Iijima,~S. Direct evidence
  for atomic defects in graphene layers. \emph{Nature} \textbf{2004},
  \emph{430}, 870--873\relax
\mciteBstWouldAddEndPuncttrue
\mciteSetBstMidEndSepPunct{\mcitedefaultmidpunct}
{\mcitedefaultendpunct}{\mcitedefaultseppunct}\relax
\EndOfBibitem
\bibitem[Meyer \latin{et~al.}(2008)Meyer, Kisielowski, Erni, Rossell, Crommie,
  and Zettl]{Meyer2008}
Meyer,~J.~C.; Kisielowski,~C.; Erni,~R.; Rossell,~M.~D.; Crommie,~M.~F.;
  Zettl,~A. Direct Imaging of Lattice Atoms and Topological Defects in Graphene
  Membranes. \emph{Nano Letters} \textbf{2008}, \emph{8}, 3582--3586\relax
\mciteBstWouldAddEndPuncttrue
\mciteSetBstMidEndSepPunct{\mcitedefaultmidpunct}
{\mcitedefaultendpunct}{\mcitedefaultseppunct}\relax
\EndOfBibitem
\bibitem[Cançado \latin{et~al.}(2011)Cançado, Jorio, Ferreira, Stavale,
  Achete, Capaz, Moutinho, Lombardo, Kulmala, and Ferrari]{Cancado2011}
Cançado,~L.~G.; Jorio,~A.; Ferreira,~E. H.~M.; Stavale,~F.; Achete,~C.~A.;
  Capaz,~R.~B.; Moutinho,~M. V.~O.; Lombardo,~A.; Kulmala,~T.~S.;
  Ferrari,~A.~C. Quantifying Defects in Graphene via Raman Spectroscopy at
  Different Excitation Energies. \emph{Nano Letters} \textbf{2011}, \emph{11},
  3190--3196\relax
\mciteBstWouldAddEndPuncttrue
\mciteSetBstMidEndSepPunct{\mcitedefaultmidpunct}
{\mcitedefaultendpunct}{\mcitedefaultseppunct}\relax
\EndOfBibitem
\bibitem[Mao \latin{et~al.}(2016)Mao, Jiang, Moldovan, Li, Watanabe, Taniguchi,
  Masir, Peeters, and Andrei]{Mao2016}
Mao,~J.; Jiang,~Y.; Moldovan,~D.; Li,~G.; Watanabe,~K.; Taniguchi,~T.;
  Masir,~M.~R.; Peeters,~F.~M.; Andrei,~E.~Y. Realization of a tunable
  artificial atom at a supercritically charged vacancy in graphene.
  \emph{Nature Physics} \textbf{2016}, \emph{12}, 545--549\relax
\mciteBstWouldAddEndPuncttrue
\mciteSetBstMidEndSepPunct{\mcitedefaultmidpunct}
{\mcitedefaultendpunct}{\mcitedefaultseppunct}\relax
\EndOfBibitem
\bibitem[Zhang \latin{et~al.}(2016)Zhang, Li, Huang, Li, Qiao, Wang, Yin, Bai,
  Duan, and He]{Zhang2016}
Zhang,~Y.; Li,~S.-Y.; Huang,~H.; Li,~W.-T.; Qiao,~J.-B.; Wang,~W.-X.;
  Yin,~L.-J.; Bai,~K.-K.; Duan,~W.; He,~L. Scanning Tunneling Microscopy of the
  π Magnetism of a Single Carbon Vacancy in Graphene. \emph{Physical Review
  Letters} \textbf{2016}, \emph{117}, 166801\relax
\mciteBstWouldAddEndPuncttrue
\mciteSetBstMidEndSepPunct{\mcitedefaultmidpunct}
{\mcitedefaultendpunct}{\mcitedefaultseppunct}\relax
\EndOfBibitem
\bibitem[Son \latin{et~al.}(2006)Son, Cohen, and Louie]{Son2006}
Son,~Y.-W.; Cohen,~M.~L.; Louie,~S.~G. Energy Gaps in Graphene Nanoribbons.
  \emph{Physical Review Letters} \textbf{2006}, \emph{97}, 216803\relax
\mciteBstWouldAddEndPuncttrue
\mciteSetBstMidEndSepPunct{\mcitedefaultmidpunct}
{\mcitedefaultendpunct}{\mcitedefaultseppunct}\relax
\EndOfBibitem
\bibitem[Banhart \latin{et~al.}(2011)Banhart, Kotakoski, and
  Krasheninnikov]{Banhart2011}
Banhart,~F.; Kotakoski,~J.; Krasheninnikov,~A.~V. Structural Defects in
  Graphene. \emph{ACS Nano} \textbf{2011}, \emph{5}, 26--41\relax
\mciteBstWouldAddEndPuncttrue
\mciteSetBstMidEndSepPunct{\mcitedefaultmidpunct}
{\mcitedefaultendpunct}{\mcitedefaultseppunct}\relax
\EndOfBibitem
\bibitem[Thiemann \latin{et~al.}(2025)Thiemann, Scalliet, Müller, and
  Michaelides]{Thiemann2025}
Thiemann,~F.~L.; Scalliet,~C.; Müller,~E.~A.; Michaelides,~A. Defects induce
  phase transition from dynamic to static rippling in graphene.
  \emph{Proceedings of the National Academy of Sciences} \textbf{2025},
  \emph{122}, e2416932122\relax
\mciteBstWouldAddEndPuncttrue
\mciteSetBstMidEndSepPunct{\mcitedefaultmidpunct}
{\mcitedefaultendpunct}{\mcitedefaultseppunct}\relax
\EndOfBibitem
\bibitem[Valencia and Caldas(2017)Valencia, and Caldas]{Valencia2017}
Valencia,~A.~M.; Caldas,~M.~J. Single vacancy defect in graphene: Insights into
  its magnetic properties from theoretical modeling. \emph{Physical Review B}
  \textbf{2017}, \emph{96}, 125431\relax
\mciteBstWouldAddEndPuncttrue
\mciteSetBstMidEndSepPunct{\mcitedefaultmidpunct}
{\mcitedefaultendpunct}{\mcitedefaultseppunct}\relax
\EndOfBibitem
\bibitem[Özçelik \latin{et~al.}(2013)Özçelik, Gurel, and
  Ciraci]{Formation2013}
Özçelik,~V.~O.; Gurel,~H.~H.; Ciraci,~S. Self-healing of vacancy defects in
  single-layer graphene and silicene. \emph{Physical Review B} \textbf{2013},
  \emph{88}, 45440\relax
\mciteBstWouldAddEndPuncttrue
\mciteSetBstMidEndSepPunct{\mcitedefaultmidpunct}
{\mcitedefaultendpunct}{\mcitedefaultseppunct}\relax
\EndOfBibitem
\bibitem[Yazyev and Helm(2007)Yazyev, and Helm]{Yazyev2007}
Yazyev,~O.~V.; Helm,~L. Defect-induced magnetism in graphene. \emph{Physical
  Review B} \textbf{2007}, \emph{75}, 125408\relax
\mciteBstWouldAddEndPuncttrue
\mciteSetBstMidEndSepPunct{\mcitedefaultmidpunct}
{\mcitedefaultendpunct}{\mcitedefaultseppunct}\relax
\EndOfBibitem
\bibitem[Palacios and Ynduráin(2012)Palacios, and Ynduráin]{Palacios2012}
Palacios,~J.~J.; Ynduráin,~F. Critical analysis of vacancy-induced magnetism
  in monolayer and bilayer graphene. \emph{Physical Review B} \textbf{2012},
  \emph{85}, 245443\relax
\mciteBstWouldAddEndPuncttrue
\mciteSetBstMidEndSepPunct{\mcitedefaultmidpunct}
{\mcitedefaultendpunct}{\mcitedefaultseppunct}\relax
\EndOfBibitem
\bibitem[Padmanabhan and Nanda(2016)Padmanabhan, and Nanda]{Padmanabhan2016}
Padmanabhan,~H.; Nanda,~B. R.~K. Intertwined lattice deformation and magnetism
  in monovacancy graphene. \emph{Physical Review B} \textbf{2016}, \emph{93},
  165403\relax
\mciteBstWouldAddEndPuncttrue
\mciteSetBstMidEndSepPunct{\mcitedefaultmidpunct}
{\mcitedefaultendpunct}{\mcitedefaultseppunct}\relax
\EndOfBibitem
\bibitem[Cabrera-Sanfelix and Darling(2007)Cabrera-Sanfelix, and
  Darling]{Cabrera-Sanfelix2007}
Cabrera-Sanfelix,~P.; Darling,~G.~R. Dissociative Adsorption of Water at
  Vacancy Defects in Graphite. \emph{The Journal of Physical Chemistry C}
  \textbf{2007}, \emph{111}, 18258--18263\relax
\mciteBstWouldAddEndPuncttrue
\mciteSetBstMidEndSepPunct{\mcitedefaultmidpunct}
{\mcitedefaultendpunct}{\mcitedefaultseppunct}\relax
\EndOfBibitem
\bibitem[Kostov \latin{et~al.}(2005)Kostov, Santiso, George, Gubbins, and
  Nardelli]{PhysRevLett.95.136105}
Kostov,~M.~K.; Santiso,~E.~E.; George,~A.~M.; Gubbins,~K.~E.; Nardelli,~M.~B.
  Dissociation of Water on Defective Carbon Substrates. \emph{Physical Review
  Letters} \textbf{2005}, \emph{95}, 136105\relax
\mciteBstWouldAddEndPuncttrue
\mciteSetBstMidEndSepPunct{\mcitedefaultmidpunct}
{\mcitedefaultendpunct}{\mcitedefaultseppunct}\relax
\EndOfBibitem
\bibitem[Liang \latin{et~al.}(2021)Liang, Li, Wang, Bu, and Zhang]{Liang2021}
Liang,~Z.; Li,~K.; Wang,~Z.; Bu,~Y.; Zhang,~J. Adsorption and reaction
  mechanisms of single and double H2O molecules on graphene surfaces with
  defects: a density functional theory study. \emph{Physical Chemistry Chemical
  Physics} \textbf{2021}, \emph{23}, 19071--19082\relax
\mciteBstWouldAddEndPuncttrue
\mciteSetBstMidEndSepPunct{\mcitedefaultmidpunct}
{\mcitedefaultendpunct}{\mcitedefaultseppunct}\relax
\EndOfBibitem
\bibitem[Campbell and Sellers(2012)Campbell, and Sellers]{Campbell2012}
Campbell,~C.~T.; Sellers,~J. R.~V. The Entropies of Adsorbed Molecules.
  \emph{Journal of the American Chemical Society} \textbf{2012}, \emph{134},
  18109--18115\relax
\mciteBstWouldAddEndPuncttrue
\mciteSetBstMidEndSepPunct{\mcitedefaultmidpunct}
{\mcitedefaultendpunct}{\mcitedefaultseppunct}\relax
\EndOfBibitem
\bibitem[Bourissou \latin{et~al.}(1999)Bourissou, Guerret, Gabbaï, and
  Bertrand]{Bourissou1999}
Bourissou,~D.; Guerret,~O.; Gabbaï,~F.~P.; Bertrand,~G. Stable Carbenes.
  \emph{Chemical Reviews} \textbf{1999}, \emph{100}, 39--92\relax
\mciteBstWouldAddEndPuncttrue
\mciteSetBstMidEndSepPunct{\mcitedefaultmidpunct}
{\mcitedefaultendpunct}{\mcitedefaultseppunct}\relax
\EndOfBibitem
\bibitem[Advincula \latin{et~al.}(2025)Advincula, Fong, Michaelides, and
  Schran]{Advincula2025}
Advincula,~X.~R.; Fong,~K.~D.; Michaelides,~A.; Schran,~C. Protons Accumulate
  at the Graphene–Water Interface. \emph{ACS Nano} \textbf{2025}, \emph{19},
  17728--17737\relax
\mciteBstWouldAddEndPuncttrue
\mciteSetBstMidEndSepPunct{\mcitedefaultmidpunct}
{\mcitedefaultendpunct}{\mcitedefaultseppunct}\relax
\EndOfBibitem
\bibitem[Batatia \latin{et~al.}(2025)Batatia, Benner, Chiang, Elena, Kovács,
  Riebesell, Advincula, Asta, Avaylon, Baldwin, Berger, Bernstein, Bhowmik,
  Bigi, Blau, Cărare, Ceriotti, Chong, Darby, De, Pia, Deringer, Elijošius,
  El-Machachi, Fako, Falcioni, Ferrari, Gardner, Gawkowski, Genreith-Schriever,
  George, Goodall, Grandel, Grey, Grigorev, Han, Handley, Heenen, Hermansson,
  Ho, Hofmann, Holm, Jaafar, Jakob, Jung, Kapil, Kaplan, Karimitari, Kermode,
  Kourtis, Kroupa, Kullgren, Kuner, Kuryla, Liepuoniute, Lin, Margraf, Magdău,
  Michaelides, Moore, Naik, Niblett, Norwood, O’Neill, Ortner, Persson,
  Reuter, Rosen, Rosset, Schaaf, Schran, Shi, Sivonxay, Stenczel, Sutton,
  Svahn, Swinburne, Tilly, van~der Oord, Vargas, Varga-Umbrich, Vegge,
  Vondrák, Wang, Witt, Wolf, Zills, and Csányi]{Batatia2023}
Batatia,~I.; Benner,~P.; Chiang,~Y.; Elena,~A.~M.; Kovács,~D.~P.;
  Riebesell,~J.; Advincula,~X.~R.; Asta,~M.; Avaylon,~M.; Baldwin,~W.~J.
  \latin{et~al.}  A foundation model for atomistic materials chemistry.
  \emph{The Journal of Chemical Physics} \textbf{2025}, \emph{163},
  184110\relax
\mciteBstWouldAddEndPuncttrue
\mciteSetBstMidEndSepPunct{\mcitedefaultmidpunct}
{\mcitedefaultendpunct}{\mcitedefaultseppunct}\relax
\EndOfBibitem
\bibitem[Blum \latin{et~al.}(2009)Blum, Gehrke, Hanke, Havu, Havu, Ren, Reuter,
  and Scheffler]{Blum2009}
Blum,~V.; Gehrke,~R.; Hanke,~F.; Havu,~P.; Havu,~V.; Ren,~X.; Reuter,~K.;
  Scheffler,~M. Ab initio molecular simulations with numeric atom-centered
  orbitals. \emph{Computer Physics Communications} \textbf{2009}, \emph{180},
  2175--2196\relax
\mciteBstWouldAddEndPuncttrue
\mciteSetBstMidEndSepPunct{\mcitedefaultmidpunct}
{\mcitedefaultendpunct}{\mcitedefaultseppunct}\relax
\EndOfBibitem
\bibitem[Ren \latin{et~al.}(2012)Ren, Rinke, Blum, Wieferink, Tkatchenko,
  Sanfilippo, Reuter, and Scheffler]{Ren2012}
Ren,~X.; Rinke,~P.; Blum,~V.; Wieferink,~J.; Tkatchenko,~A.; Sanfilippo,~A.;
  Reuter,~K.; Scheffler,~M. Resolution-of-identity approach to Hartree–Fock,
  hybrid density functionals, RPA, MP2 and GW with numeric atom-centered
  orbital basis functions. \emph{New Journal of Physics} \textbf{2012},
  \emph{14}, 53020\relax
\mciteBstWouldAddEndPuncttrue
\mciteSetBstMidEndSepPunct{\mcitedefaultmidpunct}
{\mcitedefaultendpunct}{\mcitedefaultseppunct}\relax
\EndOfBibitem
\bibitem[Lehtola \latin{et~al.}(2018)Lehtola, Steigemann, Oliveira, and
  Marques]{Lehtola2018}
Lehtola,~S.; Steigemann,~C.; Oliveira,~M. J.~T.; Marques,~M. A.~L. Recent
  developments in libxc — A comprehensive library of functionals for density
  functional theory. \emph{SoftwareX} \textbf{2018}, \emph{7}, 1--5\relax
\mciteBstWouldAddEndPuncttrue
\mciteSetBstMidEndSepPunct{\mcitedefaultmidpunct}
{\mcitedefaultendpunct}{\mcitedefaultseppunct}\relax
\EndOfBibitem
\bibitem[Perdew \latin{et~al.}(1996)Perdew, Burke, and
  Ernzerhof]{PhysRevLett.77.3865}
Perdew,~J.~P.; Burke,~K.; Ernzerhof,~M. Generalized Gradient Approximation Made
  Simple. \emph{Physical Review Letters} \textbf{1996}, \emph{77},
  3865--3868\relax
\mciteBstWouldAddEndPuncttrue
\mciteSetBstMidEndSepPunct{\mcitedefaultmidpunct}
{\mcitedefaultendpunct}{\mcitedefaultseppunct}\relax
\EndOfBibitem
\bibitem[Zhang and Yang(1998)Zhang, and Yang]{PhysRevLett.80.890}
Zhang,~Y.; Yang,~W. Comment on ``Generalized Gradient Approximation Made
  Simple''. \emph{Physical Review Letters} \textbf{1998}, \emph{80}, 890\relax
\mciteBstWouldAddEndPuncttrue
\mciteSetBstMidEndSepPunct{\mcitedefaultmidpunct}
{\mcitedefaultendpunct}{\mcitedefaultseppunct}\relax
\EndOfBibitem
\bibitem[Adamo and Barone(1999)Adamo, and Barone]{doi:10.1063/1.478522}
Adamo,~C.; Barone,~V. Toward reliable density functional methods without
  adjustable parameters: The PBE0 model. \emph{The Journal of Chemical Physics}
  \textbf{1999}, \emph{110}, 6158--6170\relax
\mciteBstWouldAddEndPuncttrue
\mciteSetBstMidEndSepPunct{\mcitedefaultmidpunct}
{\mcitedefaultendpunct}{\mcitedefaultseppunct}\relax
\EndOfBibitem
\bibitem[Grimme \latin{et~al.}(2010)Grimme, Antony, Ehrlich, and
  Krieg]{doi:10.1063/1.3382344}
Grimme,~S.; Antony,~J.; Ehrlich,~S.; Krieg,~H. A consistent and accurate ab
  initio parametrization of density functional dispersion correction (DFT-D)
  for the 94 elements H-Pu. \emph{The Journal of Chemical Physics}
  \textbf{2010}, \emph{132}, 154104\relax
\mciteBstWouldAddEndPuncttrue
\mciteSetBstMidEndSepPunct{\mcitedefaultmidpunct}
{\mcitedefaultendpunct}{\mcitedefaultseppunct}\relax
\EndOfBibitem
\bibitem[Thompson \latin{et~al.}(2022)Thompson, Aktulga, Berger, Bolintineanu,
  Brown, Crozier, in~'t Veld, Kohlmeyer, Moore, Nguyen, Shan, Stevens,
  Tranchida, Trott, and Plimpton]{Thompson2022}
Thompson,~A.~P.; Aktulga,~H.~M.; Berger,~R.; Bolintineanu,~D.~S.; Brown,~W.~M.;
  Crozier,~P.~S.; in~'t Veld,~P.~J.; Kohlmeyer,~A.; Moore,~S.~G.; Nguyen,~T.~D.
  \latin{et~al.}  LAMMPS - a flexible simulation tool for particle-based
  materials modeling at the atomic, meso, and continuum scales. \emph{Computer
  Physics Communications} \textbf{2022}, \emph{271}, 108171\relax
\mciteBstWouldAddEndPuncttrue
\mciteSetBstMidEndSepPunct{\mcitedefaultmidpunct}
{\mcitedefaultendpunct}{\mcitedefaultseppunct}\relax
\EndOfBibitem
\bibitem[Barducci \latin{et~al.}(2008)Barducci, Bussi, and
  Parrinello]{Barducci2008}
Barducci,~A.; Bussi,~G.; Parrinello,~M. Well-Tempered Metadynamics: A Smoothly
  Converging and Tunable Free-Energy Method. \emph{Physical Review Letters}
  \textbf{2008}, \emph{100}, 20603\relax
\mciteBstWouldAddEndPuncttrue
\mciteSetBstMidEndSepPunct{\mcitedefaultmidpunct}
{\mcitedefaultendpunct}{\mcitedefaultseppunct}\relax
\EndOfBibitem
\bibitem[Raiteri \latin{et~al.}(2006)Raiteri, Laio, Gervasio, Micheletti, and
  Parrinello]{Raiteri2006}
Raiteri,~P.; Laio,~A.; Gervasio,~F.~L.; Micheletti,~C.; Parrinello,~M.
  Efficient Reconstruction of Complex Free Energy Landscapes by Multiple
  Walkers Metadynamics. \emph{The Journal of Physical Chemistry B}
  \textbf{2006}, \emph{110}, 3533--3539\relax
\mciteBstWouldAddEndPuncttrue
\mciteSetBstMidEndSepPunct{\mcitedefaultmidpunct}
{\mcitedefaultendpunct}{\mcitedefaultseppunct}\relax
\EndOfBibitem
\bibitem[Bonomi \latin{et~al.}(2009)Bonomi, Branduardi, Bussi, Camilloni,
  Provasi, Raiteri, Donadio, Marinelli, Pietrucci, Broglia, and
  Parrinello]{PLUMED}
Bonomi,~M.; Branduardi,~D.; Bussi,~G.; Camilloni,~C.; Provasi,~D.; Raiteri,~P.;
  Donadio,~D.; Marinelli,~F.; Pietrucci,~F.; Broglia,~R.~A. \latin{et~al.}
  PLUMED: A portable plugin for free-energy calculations with molecular
  dynamics. \emph{Computer Physics Communications} \textbf{2009}, \emph{180},
  1961--1972\relax
\mciteBstWouldAddEndPuncttrue
\mciteSetBstMidEndSepPunct{\mcitedefaultmidpunct}
{\mcitedefaultendpunct}{\mcitedefaultseppunct}\relax
\EndOfBibitem
\bibitem[Tribello \latin{et~al.}(2014)Tribello, Bonomi, Branduardi, Camilloni,
  and Bussi]{PLUMED2}
Tribello,~G.~A.; Bonomi,~M.; Branduardi,~D.; Camilloni,~C.; Bussi,~G. PLUMED 2:
  New feathers for an old bird. \emph{Computer Physics Communications}
  \textbf{2014}, \emph{185}, 604--613\relax
\mciteBstWouldAddEndPuncttrue
\mciteSetBstMidEndSepPunct{\mcitedefaultmidpunct}
{\mcitedefaultendpunct}{\mcitedefaultseppunct}\relax
\EndOfBibitem
\end{mcitethebibliography}
\providecommand{\latin}[1]{#1}
\makeatletter
\providecommand{\doi}
  {\begingroup\let\do\@makeother\dospecials
  \catcode`\{=1 \catcode`\}=2 \doi@aux}
\providecommand{\doi@aux}[1]{\endgroup\texttt{#1}}
\makeatother
\providecommand*\mcitethebibliography{\thebibliography}
\csname @ifundefined\endcsname{endmcitethebibliography}
  {\let\endmcitethebibliography\endthebibliography}{}

\end{document}


\maketitle

\tableofcontents

\section{Model development}

\vspace{25mm}

\begin{figure*}[h!]
\centering
    \includegraphics[width=1.00\textwidth]{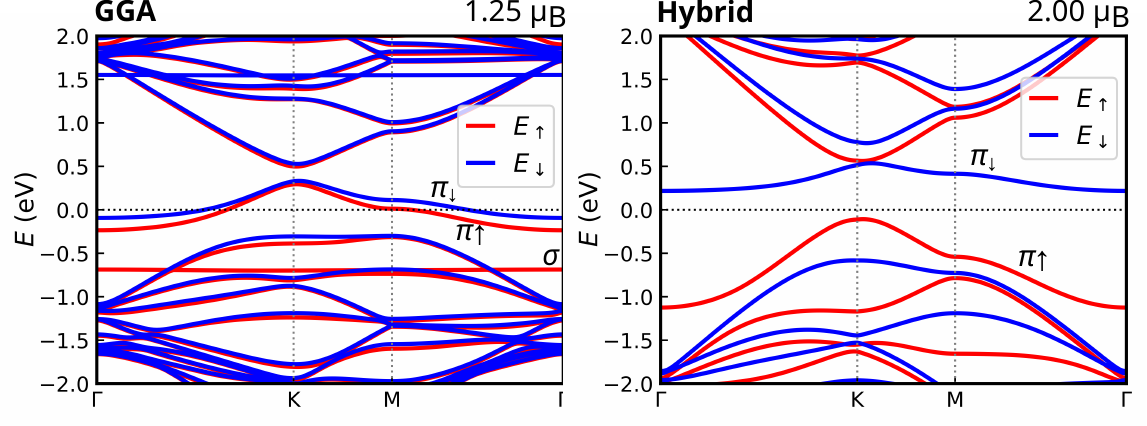}
\caption{Band structure of the graphene single vacancy calculated using GGA (left) and hybrid (right) DFT levels of theory.
We perform calculations on a $7 \times 7$ hexagonal cell with a single graphene defect. 
GGA calculations are performed using revPBE-D3 and the hybrid calculations with revPBE0-D3.
The characteristic $\sigma$ and $\pi$ bands are highlighted on the figures (in the case of the hybrid band structure, the $\sigma$ resides at -2.4 eV). 
}
\label{fig:band_structure}
\end{figure*}

\begin{figure*}[h!]
\centering
    \includegraphics[width=0.90\textwidth]{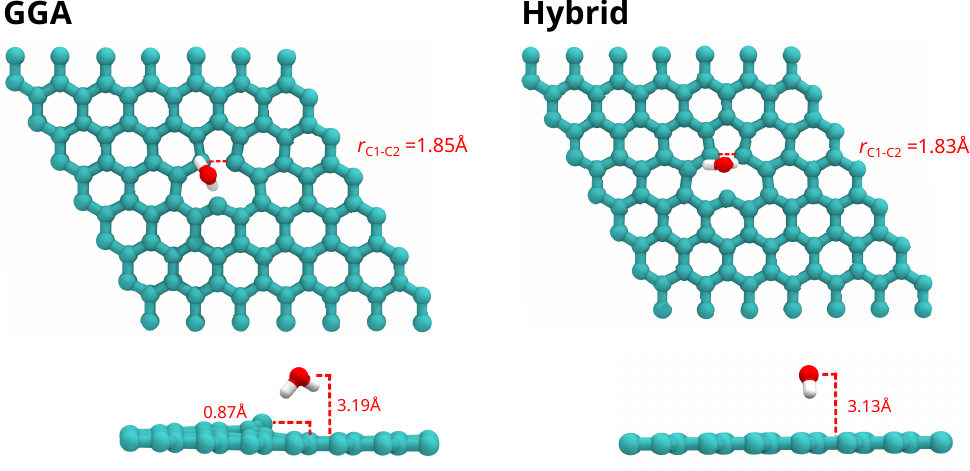}
\caption{Geometry optimized structures of the graphene SV + 1\ch{H2O} system using MACE potentials trained at GGA and hybrid DFT levels of theory. 
The GGA MACE potential was trained on revPBE-D3-labeled data, and the hybrid model on revPBE0-D3 data.
In the presence of a single water molecule, the GGA SV undergoes puckering out of the plane of the sheet; no puckering is observed in the case of the hybrid SV. 
}
\label{fig:gga_hybrid_struct}
\end{figure*}

\begin{figure}[h!]
\centering
    \includegraphics[width=0.65\textwidth]{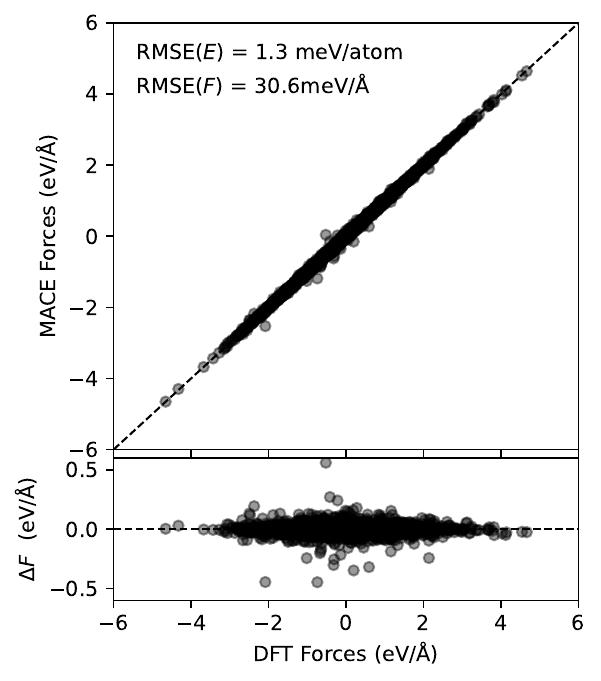}
\caption{Comparison of MACE and DFT (revPBE0-D3) force predictions.  
Forces are computed for $\sim$ 200 structures extracted from the test dataset. 
A random subsample of 5000 forces is compared between MACE and DFT methodologies. 
}
\label{fig:frc_err}
\end{figure}

\vspace{20mm}

\clearpage

\section{NEB predictions}

\vspace{50mm}

\begin{figure*}[h!]
\centering
    \includegraphics[width=1.0\textwidth]{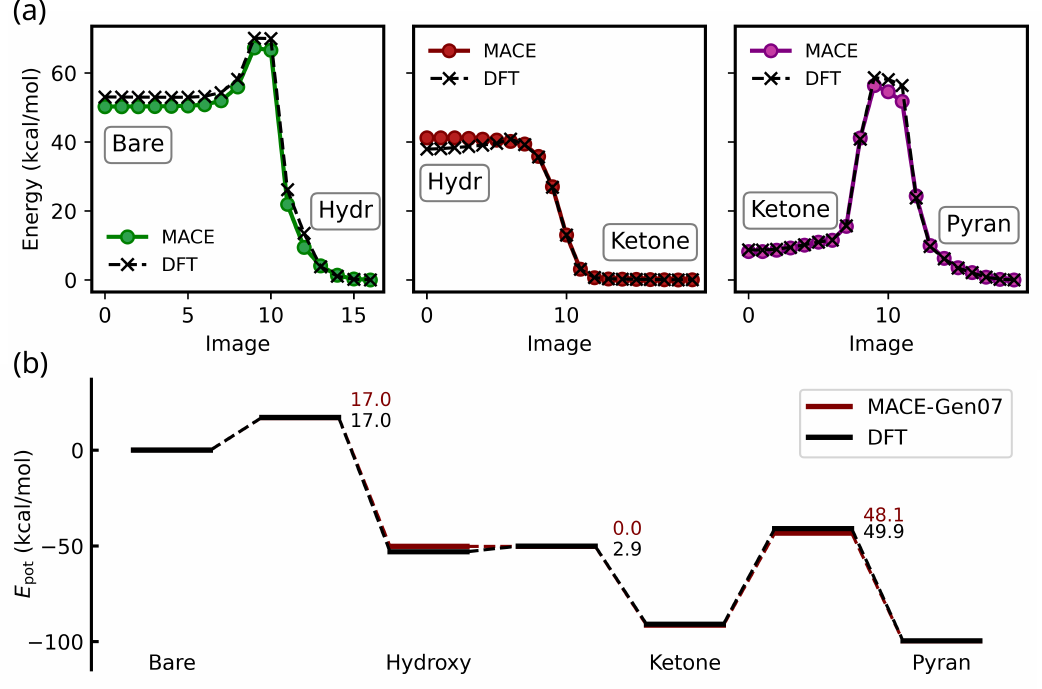}
\caption{Potential energies for the chemisorption of a single water molecule at the graphene single vacancy. 
\textbf{(a)} Nudged elastic band profiles for each of the key decomposition processes: the bare vacancy to Hydroxy-SV, the Hydroxy-SV to SV-ketone, and the SV-Ketone to SV-Pyran. 
Images are generated using a MACE potential trained at the hybrid DFT level of theory. 
MACE potential energies are plotted alongside revPBE0-D3 DFT energies obtained from single-point calculations of each image. 
\textbf{(b)} State energy profile connecting the bare single vacancy to the Pyran product state. 
Transition state energies are shown for both MACE and DFT calculations. 
}
\label{fig:nebs}
\end{figure*}

\clearpage

\section{Pyran Formation}

\vspace{50mm}

\begin{figure*}[h!]
\centering
    \includegraphics[width=1.0\textwidth]{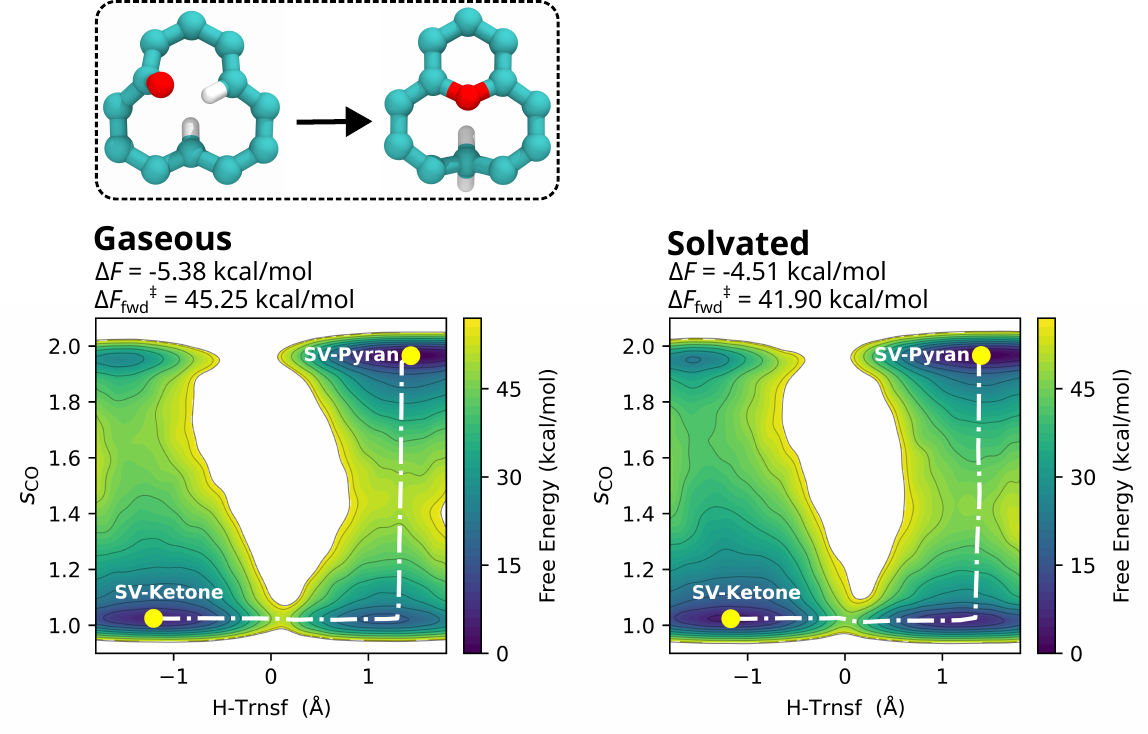}
\caption{SV-Pyran formation is similar under gaseous and solvated conditions. 
Free energy surfaces for the conversion of the SV-Ketone defect to the SV-Pyran defect are shown as a function of two collective variables: a proton-transfer coordinate (x-axis) and a C-O coordination number (y-axis) for the defect carbons. 
Minimum energy pathways are highlighted on the free energy surfaces, with free energy differences ($\Delta F$) and barrier heights ($\Delta F ^{\ddag}$) reported above each plot. 
}
\label{fig:pyran}
\end{figure*}

\clearpage

%

%

%
%
%
%
%
%
%
%
%
%
%
%
%
%
%
%
%
%
%
%
%
%
